\documentclass[conference]{IEEEtran}
\IEEEoverridecommandlockouts

\usepackage{bm}
\usepackage{cite}
\usepackage{amsmath,amssymb,amsfonts}
\usepackage{algorithmic}
\usepackage{graphicx}
\usepackage{textcomp}
\usepackage{xcolor}
\def\BibTeX{{\rm B\kern-.05em{\sc i\kern-.025em b}\kern-.08em
    T\kern-.1667em\lower.7ex\hbox{E}\kern-.125emX}}
\usepackage{multirow}

\usepackage{url}
\usepackage{orcidlink}

\begin{document}

\title{EConv-TasNet: Efficient Conv-TasNet for Effective Speech Separation}

\author{Pei-Chun Chang \\
\IEEEauthorblockA{\textit{Novatek Microelectronics Corporation} \\
Hsinchu, Taiwan \\
pei\_chang@novatek.com.tw \orcidlink{0000-0000-0000-0000}}
\and
\IEEEauthorblockN{Chuan-Yi Liu}
\IEEEauthorblockA{\textit{Novatek Microelectronics Corporation} \\
Hsinchu, Taiwan \\
daniel\_liu@novatek.com.tw}

}

\maketitle

\begin{abstract}
Conv-TasNet has served as a strong baseline for time-domain speech separation, and many studies have extended it with advanced architectures such as dual-path networks, U-Nets, and attention mechanisms. 
However, these methods often introduce high computational cost and complexity, limiting their deployment in resource-constrained scenarios. 
To address this issue, we propose eConv-TasNet, an efficient variant of Conv-TasNet that improves both effectiveness and efficiency without relying on resource-intensive modules. 
The proposed model consists of a group-wise early-splitting (GES) module and a multi-group feature aggregation (MGFA) module. 
GES generates discriminative speaker embeddings at intermediate stages, while MGFA progressively aggregates these group-level representations for refined mask estimation. 
Experimental results show that eConv-TasNet reduces model size by 22.4\%, accelerates inference by 18.9\%, and improves SI-SNRi by 14.0\%–28.0\% across three public benchmarks. 
Moreover, it achieves competitive performance compared with state-of-the-art methods while requiring significantly fewer parameters and lower inference cost. 
These results demonstrate a favorable efficiency–effectiveness trade-off for edge deployment.

\end{abstract}

\begin{IEEEkeywords}
Speech separation, Conv-TasNet, low-cost neural network, early-splitting, feature aggregation.
\end{IEEEkeywords}

\section{Introduction}\label{sec:introduction}
Speech separation is a crucial task in speech processing that aims to extract individual speakers' voice sources from a single mixed signal~\cite{wang2018supervised,agrawal2023review}.
Traditional approaches predominantly rely on time–frequency (TF) domain methods~\cite{hershey2016deep,wang2023tf,xu2025tiger}.
These studies apply the short-time Fourier transform (STFT) and its inverse (iSTFT) to convert signals between the time domain and the time–frequency (TF) domain, and develop a separator to distinguish the characteristics of various sound sources.

However, STFT/iSTFT have inherent limitations, such as time-frequency resolution constrained by the non-learnable, fixed window function/size and stride settings, which result in a lack of abundance and diversity of feature representations in the separation task and challenge real-time and low-latency practice scenarios~\cite{li2022skim}.
Moreover, these works focus on magnitude estimation while reusing the mixture phase during reconstruction, which often leads to imperfect phase recovery and degraded separation quality~\cite{yang2023monaural}.
Although phase-aware or complex-domain approaches have been proposed to complement phase-informative content, they typically incur additional computational cost, which limits their applicability on edge devices~\cite{wang2018end,li2018cbldnn}.

\begin{figure}
    \centering
    \includegraphics[clip, trim= 1.5cm 12.9cm 6.6cm 3.5cm, width=\linewidth]{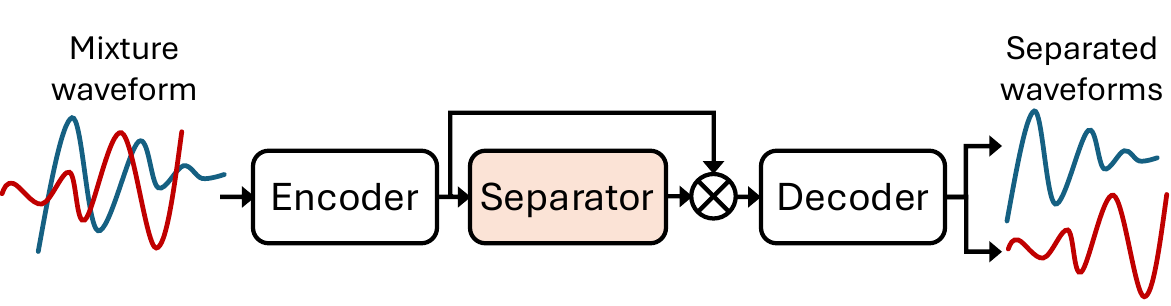}
    \caption{The TasNet pipeline for speech separation.}
    \label{fig:tasnet}
\end{figure}

TasNet~\cite{luo2018tasnet}, consisting of the encoder, separator, and decoder, is a classical solution to operate speech separation directly on the raw audio waveform rather than TF representation, as shown in Fig.~\ref{fig:tasnet}.
Their encoder–decoder architecture, implemented with convolutional/deconvolutional layers, transforms signals between the time domain and the latent space, and the separator estimates feature masks for each individual speaker.
Specifically, the variant of TasNet, Conv-TasNet~\cite{luo2019conv}, applies a fully convolutional network (FCN) pipeline to propose a separator composed of stacked temporal convolutional network (TCN) units to capture discriminative features with diverse receptive fields.
Compared to the original TasNet, Conv-TasNet achieves significant improvements in accuracy, inference speed, and computational cost.

Based on the TasNet pipeline, an encoder–decoder architecture with a separator, various approaches have been proposed to improve representation discriminability by specific techniques such as dual-path networks, multi-scale models, attention mechanisms, and recurrent strategies~\cite{li2025advances}.
These mechanisms leverage the extraction of abundant acoustic features to improve separation accuracy, especially speech quality.
Dual-path networks aim to collaborate intra- and inter-chunk information~\cite{luo2020dual, subakan2021attention} to sequentially model local and global information.
Multi-scale models typically employ hierarchical architectures, such as U-Net, to aggregate multi-resolution features and capture diverse features~\cite{tzinis2020sudo, li2023efficient}.
Attention mechanisms (e.g., Transformers) are adopted to flexibly capture long-range temporal dependencies and emphasize informative features, thereby improving robustness and separation~\cite{lam2021effective, zhao2023mossformer, zhao2024mossformer2}.
Modeling long-term dependency via recurrent strategies can be used to progressively aggregate contextual information across temporal/frequency aspects~\cite{lam2021sandglasset, mu2023multi}.


\begin{figure*}[t]
    \centering
    \includegraphics[clip, trim= 0.2cm 6.3cm 9.2cm 6cm, width=0.92\linewidth]{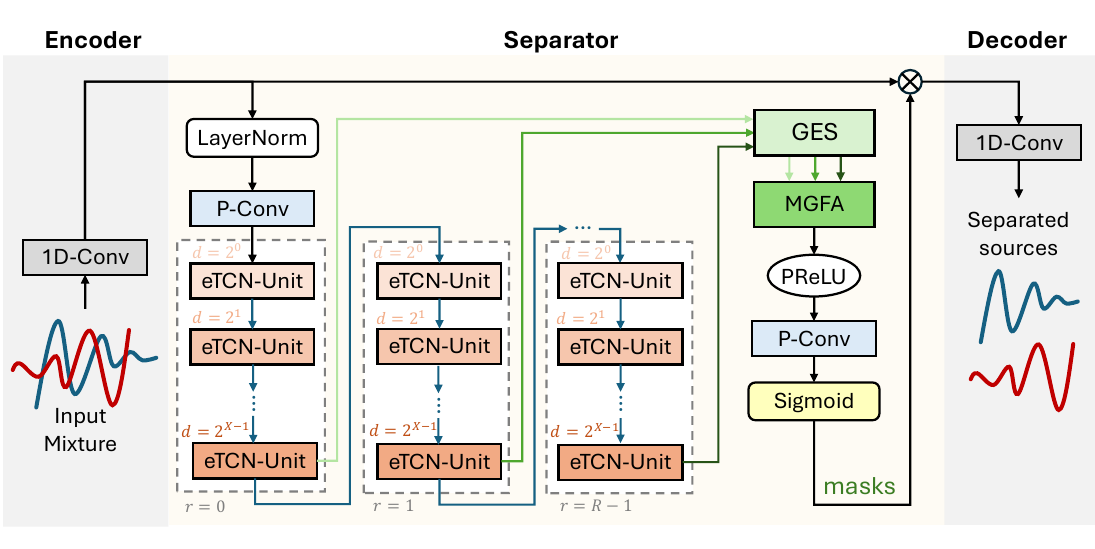}
    \caption{The proposed eConv-TasNet architecture for speech separation.}
    \label{fig:overall}
\end{figure*}

Furthermore, hybrid architectures that combine these techniques, such as recurrent U-Nets~\cite{tzinis2020}, dual-path/U-Net models with Transformers~\cite{shin2024separate}, recurrent attention~\cite{yang2023monaural}, have also been explored to improve separation quality and robustness.
However, compared with Conv-TasNet, these approaches generally incur substantial memory consumption and high computational complexity during both training and inference, which remains a critical challenge to deployment in real-world acoustic scenarios on resource-constrained edge devices.
Motivated by this limitation, we aim to design a new fully convolutional architecture based on Conv-TasNet to improve the efficiency and effectiveness of speech separation while avoiding high computational costs and inference time.

In this study, we propose an efficient variant of Conv-TasNet, termed eConv-TasNet, which enhances both effectiveness and efficiency for speech separation.
By introducing group-wise early-splitting (GES) and multi-group feature aggregation (MGFA) modules, speech separation performance can be significantly improved while reducing computational cost and inference time.
To sum up, the contributions of this study are:
\begin{itemize}
    \item We have proposed eConv-TasNet, consisting of GES and MGFA modules, as an efficient extension of Conv-TasNet for speech separation.
    \item The proposed GES extracts speaker-dependent representations at the group level, thereby reducing computational cost while increasing feature discriminability.
    \item The MGFA module progressively refines and aggregates group-level speaker representations into compact embeddings.
    \item Our extensive experiments have demonstrated that our eConv-TasNet outperforms the original Conv-TasNet in separation accuracy and inference speed, and show potential for practical application on edge devices compared with state-of-the-art methods on WSJ0-2mix, WHAM!, and Libri2Mix datasets.
\end{itemize}

\section{The proposed method}\label{sec:method}

\subsection{Overview}
In this study, we propose an efficient variant of Conv-TasNet~\cite{luo2019conv}, termed eConv-TasNet, to balance performance and computational cost.
As shown in Fig.~\ref{fig:overall}, the proposed architecture retains the original encoder-decoder framework and introduces group-wise early-splitting (GES) and multi-group feature aggregation (MGFA) modules in the separator to improve the discriminability of speaker-related representations.
These designs leverage progressively refined speaker embeddings from fewer representations, significantly reducing computational cost and model complexity, and resulting in more effective and efficient separation performance.

\subsection{The limitation of Conv-TasNet}
Conv-TasNet employs a TCN-based separator consisting of several TCN-units with repeated dilation settings to estimate a feature mask for each speaker. 
In the original configuration, the separator is formed by stacking $X$ TCN-units and repeating this $R$ times. 
Each unit produces two representations, denoted as $\bm{F}$ and $\bm{O}$, where $\bm{F}$ contributes to mask estimation and $\bm{O}$ is forwarded to the next unit. 
To obtain the semantic embeddings for mask estimation, Conv-TasNet directly aggregates $X \times R$ representations from all unit outputs $\bm{F}$ via summation operation.
However, this approach inevitably brings computational redundancy and inefficiency, especially in real-time or resource-constrained scenarios.

In practice, the representations calculated from units in the separator make a similar contribution to near units.
In other words, aggregating representations from all units may accumulate several redundancies, which may affect the information efficiency and diversity.
Hence, this study aims to first extract speaker-related information from each group of units rather than the last accumulated features, and further progressively aggregate them to refine speaker embeddings rather than directly summing.



\begin{figure}[t]
    \centering
    \includegraphics[clip, trim= 0.1cm 12cm 18.4cm 4.2cm, width=\linewidth]{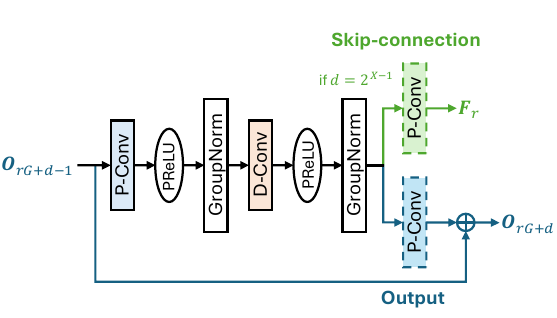}
    \caption{The illustration of the eTCN-Unit module in eConv-TasNet.}
    \label{fig:eTCN}
\end{figure}

\subsection{Efficient Conv-TasNet (eConv-TasNet)}
Followed by the original Conv-TasNet, we consider $X$ eTCN-units as a group, where each unit adopts different dilation settings, and we then repeat the group $R$ times, as shown in Fig.~\ref{fig:overall}.
To enhance the effectiveness and efficiency of speech separation, GES is first applied to separate speaker representations from each group output, and MGFA is then utilized to aggregate all speaker representations for mask estimation.
For more details, the skip-connection path is applied only to the last eTCN-unit in a group, reducing the number of learnable parameters and computational cost compared to the original Conv-TasNet, as shown in Fig.~\ref{fig:eTCN}.
As in the original pipeline, the speaker masks are computed using a PReLU, a point-wise convolution, and a Sigmoid function applied to embeddings.
The masks are applied to mixture features from the encoder to yield individual speaker features, and the decoder is further used to produce the separated waveforms.

\subsubsection{Group-wise early-splitting (GES)}
\begin{figure}[t]
    \centering
    \includegraphics[clip, trim= 7.1cm 14.4cm 12.3cm 4.5cm, width=\linewidth]{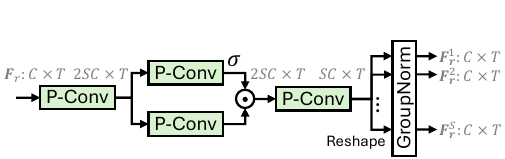}
    \caption{The illustration of the GES module.}
    \label{fig:split}
\end{figure}

The early-splitting (ES) strategy aims to obtain speaker-dependent representations earlier in the network, rather than deferring separation until the final layer of the separator~\cite{shin2024separate}.
To further improve efficiency, the proposed GES splits only the final output representation of the group formed by stacked eTCN-units, rather than splitting each individual unit.
This design not only reduces computational cost and model complexity but also preserves the most informative group-level features for subsequent processing. 
As shown in Fig.~\ref{fig:split}, our GES module is implemented using a gated linear unit (GLU), followed by a point-wise convolution and group normalization.
These $s$-th speaker-dependent representations from all groups $\bm{F}_{r}^{s}, r = 1, 2, ..., R$ are then forwarded to the MGFA module for aggregation and refinement.


\subsubsection{Multi-group feature aggregation (MGFA)}
To aggregate speaker representations across groups, we adopt an exponentially weighted moving average (EWMA) to progressively refine speaker-dependent embeddings for mask estimation while preserving information from previous groups. 
This approach adaptively balances the contributions of the current stage with the accumulated previous groups, ensuring both stability and adaptiveness in embedding learning.

Mathematically, the $r$-th embedding $\bm{E}_{r}^{s}$ for the $s$-th speaker is updated from the previous embedding $\bm{E}_{r-1}^{s}$ and current speaker representation $\bm{F}_{r}^{s}$ with a weight decay coefficient $\beta$ as follows:
\begin{equation}
\bm{E}_{r}^{s} = 
\left\{
    \begin{array}{ll}
    \bm{F}_{r}^{s}, & r=0,\\
    \beta \bm{F}_{r}^{s} + (1 - \beta)\bm{E}_{r-1}^{s}, & \text{otherwise}, 
    \end{array}
\right.
\end{equation}
where $r = 1, 2, \ldots, R$ stands for the stage index stands and $R$ represents total number of repeats in the separator. 
$\bm{F}_{r}^{s}$ denotes the group-level representation of the $s$-th speaker at the $r$-th repeat, and $\beta$ serves as a coefficient that balances the contribution between the current and previous refined embeddings.
Finally, the last embedding $\bm{E}_{R}^{s}$ is used to estimate the mask of the $s$-th speaker.



\section{Experiment setup}\label{sec:setup}

\begin{table*}[t]
    \centering
    \caption{The ablation studies on the WSJ0-2mix dataset.}
    \begin{tabular}{lccccccc}
    \hline
     Methods & Speaker-split module & MGFA & Weight-decay coefficient ($\beta$) & Dynamic mixing & \#Params (M) & SI-SNRi & RTF \\\hline
     \multirow{3}{*}{Conv-TasNet~\cite{luo2019conv}} & - & - & - & - & 4.9 & 15.3 & 0.16\\
      & - & - & - & \checkmark & 4.9 & 16.0 & 0.16 \\
      & ES~\cite{shin2024separate} & - & - & \checkmark & 5.2 & 16.4 & 0.27\\\hline
     \multirow{5}{*}{eConv-TasNet (ours)} & GES & - & - & \checkmark & 3.8 & 17.7 & 0.13\\
     & GES & \checkmark & \checkmark & 0.2 & 3.8 & 18.4 & 0.13\\
     & GES & \checkmark & \checkmark & 0.4 & 3.8 & 18.2 & 0.13\\
     & GES & \checkmark & \checkmark & 0.6 & 3.8 & 18.1 & 0.13\\
     & GES & \checkmark & \checkmark & 0.8 & 3.8 & 18.1 & 0.13\\
     \hline
    \end{tabular}
    \label{tab:ablation}
\end{table*}

\subsection{Datasets}
In this study, we evaluated our method on three popular datasets for two-speaker speech separation: WSJ0-2mix~\cite{hershey2016deep}, WHAM!~\cite{wichern2019wham}, and Libri2Mix~\cite{cosentino2020librimix}.
To conduct fair experiments and ensure model generalization, test-set mixtures were generated using speakers not included in the training phases.
For these datasets, all instances were resampled to 8 kHz, and random 4-second segments were used during the training phase, whereas full-waveform instances of varying lengths were used during the validation and evaluation phases.

The WSJ0-2mix dataset contains 30, 10, and 5 hours of training, validation, and testing data, respectively.
The training set is collected from the si\_tr\_s set in the Wall Street Journal (WSJ0)\footnote{\url{https://catalog.ldc.upenn.edu/LDC93S6A}}~\cite{Garofolo1993}.
Then, the validation and testing sets are collected from the si\_dt\_05 and si\_et\_05 sets, respectively.
Each mixture is synthesized from the speech signals of two different speakers by a relative signal-to-noise ratio (SNR) sampled from [-5, 5] dB.

The WHAM! dataset is a noisy version of the WSJ0-2mix dataset.
In this dataset, noise recordings were collected from various scenes, including cafes, restaurants, bars, etc.
Noise sounds were applied to yield a noisy mixture with SNRs uniformly sampled from [-6, 3] dB, making the mixture more challenging than the original WSJ0-2mix dataset.

The Libri2Mix dataset contains 58, 11, and 11 hours of training, validation, and testing data, respectively.
In this dataset, the speech signals are collected by randomly sampling from the train-100 set of the LibriSpeech dataset\footnote{\url{https://www.openslr.org/12}}~\cite{panayotov2015librispeech} to accelerate training, and the noise sounds were collected from WHAM! dataset.
Each mixture signal was generated by two different speakers and was adjusted in amplitude by uniformly sampling perceptually-aware Loudness Unit relative to Full Scale (LUFS)~\cite{bs1770algorithms} between -25 and -33 dB.
In this study, we used the clean version without noise sounds to conduct experiment comparisons with previous studies.

\subsection{Baseline setup}\label{sec:baseline}
In this study, we have adopted Conv-TasNet\footnote{\url{https://docs.pytorch.org/audio/main/generated/torchaudio.models.ConvTasNet.html}} as our baseline model to investigate the effectiveness and efficiency of the proposed method.
In original settings~\cite{luo2019conv}, the mixture feature is extracted by the encoder consisting of $N = 256$ filters with $L = 16$ coefficients.
The separator consists of $X \times R$ TCN-units and uses a sigmoid activation function to normalize each speaker mask, where $X$ and $R$ are set to 8 and 3, respectively.
The representation and embedding dimension is $B/Sc = 128$, and $H = 512$ for hidden layers in the eTCN-unit with kernel size of $P = 3$. 

\subsection{Implementation details}
To investigate the effectiveness and efficiency of the proposed method, our eConv-TasNet architecture used the same hyperparameters as the original Conv-TasNet, as described in Section \ref{sec:baseline}, i.e., $N=512, L=16, B/Sc=128, H=512, P=3, X=8, R=3$.
The weight decay coefficient $\beta$ in the MGFA module is set to 0.2 to progressively refine speaker embeddings.

In this study, the networks are trained for up to 200 epochs with an initial learning rate of $lr = 1e^{-3}$ using the AdamW optimizer~\cite{loshchilov2017decoupled} with a weight decay of $1e^{-2}$.
In the first epoch, we use a warm-up training scheduler for the first 1,000 steps, and then $lr$ is decayed by a factor of 0.8 when the validation loss does not improve for three consecutive epochs.
During the training phase, we apply the permutation invariant training (PIT)~\cite{kolbaek2017multitalker} strategy to avoid the effects of speaker permutation, and use negative scale-invariant signal-to-noise ratio (SI-SNR) as our training objective, in which SI-SNRi can be formulated as:
\begin{equation}
    \left\{ \begin{array}{l} 
        \mathcal{L}(\bm{S}, \hat{\bm{S}}) = 10 \log_{10} \frac{||\bm{\gamma}\bm{S}||_{2}}{||\bm{\gamma}\bm{S}-\hat{\bm{S}}||_{2}}, \\
        \bm{\gamma} = \frac{{\hat{\bm{S}}}^{\text{T}}\bm{S}}{||\bm{S}||^2_2},
        \end{array}\right.
\end{equation}
where $||\cdot||_2$ denotes $L2$-norm.
In addition, dynamic mixing~\cite{zeghidour2021wavesplit} is used for data augmentation to increase data diversity during training.

\section{Experimental results}\label{sec:result}
In this study, we used the SI-SNR improvement (SI-SNRi)~\cite{le2019sdr} and signal-to-distortion ratio improvement (SDRi)~\cite{vincent2006performance} as metrics to evaluate the speech separation performance.
To investigate model efficiency, we report the inference cost by the real-time factor (RTF) on a single thread in the 13th Gen Intel (R) Core (TM) i5-1335U 1.30 GHz CPU to evaluate the feasibility of real-time applications without GPU resources.
In addition, MAC values are calculated by inferring a 16,000-sample mixture signal to investigate the computational costs.


\subsection{Ablation studies}
We first investigate the individual contributions of the proposed GES and MGFA modules on the WSJ0-2mix dataset.
The results are summarized in Table~\ref{tab:ablation}.
We reimplemented Conv-TasNet experiments on the WSJ0-2mix dataset and obtained 15.3 dB SI-SNRi.
For a fair comparison, we applied dynamic mixing for data augmentation, improving SI-SNRi from 15.3 to 16.0 dB without increasing model size or inference cost.
When we apply the early-splitting (ES) strategy proposed in~\cite{shin2024separate} to Conv-TasNet with dynamic mixing, it further improves SI-SNRi to 16.4 dB.
However, this improvement comes at the expense of increased model complexity, resulting in a larger model size and higher inference cost.

In contrast, replacing ES with the proposed GES achieves both higher effectiveness and better efficiency.
Specifically, GES improves SI-SNRi to 17.7 dB while simultaneously reducing the model size to 3.8M parameters and lowering the RTF to 0.13.
Compared with ES, GES provides a gain of 1.3 dB in SI-SNRi while reducing the parameter count and inference cost.
These results demonstrate that introducing speaker-aware processing at the group level can generate more discriminative speaker representations while avoiding the redundancy and computational overhead introduced by previous early-splitting designs.
Furthermore, we evaluate the effectiveness of the proposed MGFA module by progressively aggregating group-level speaker embeddings. 
Integrating MGFA consistently improves separation performance across all settings, indicating that the aggregation mechanism effectively captures complementary speaker information from different groups and refines the speaker representations used for mask estimation.

Among all configurations, the best performance is achieved when the weight-decay coefficient is set to $\beta=0.2$, yielding an SI-SNRi of 18.4 dB.
This corresponds to an additional gain of 0.7 dB over the GES-only model.
That is, all MGFA variants consistently outperform the GES-only configuration, confirming the robustness of the proposed aggregation strategy.
Overall, the joint use of GES and MGFA achieves the best balance between separation quality and computational efficiency.
Compared with the Conv-TasNet, the proposed eConv-TasNet improves SI-SNRi by 12.2\% while reducing the model size and RTF by 22.4\% and 18.9\%, respectively.

\begin{figure}[t]
    \centering
    \includegraphics[clip, trim= 0.5cm 0cm 1.5cm 1cm, width=\linewidth]{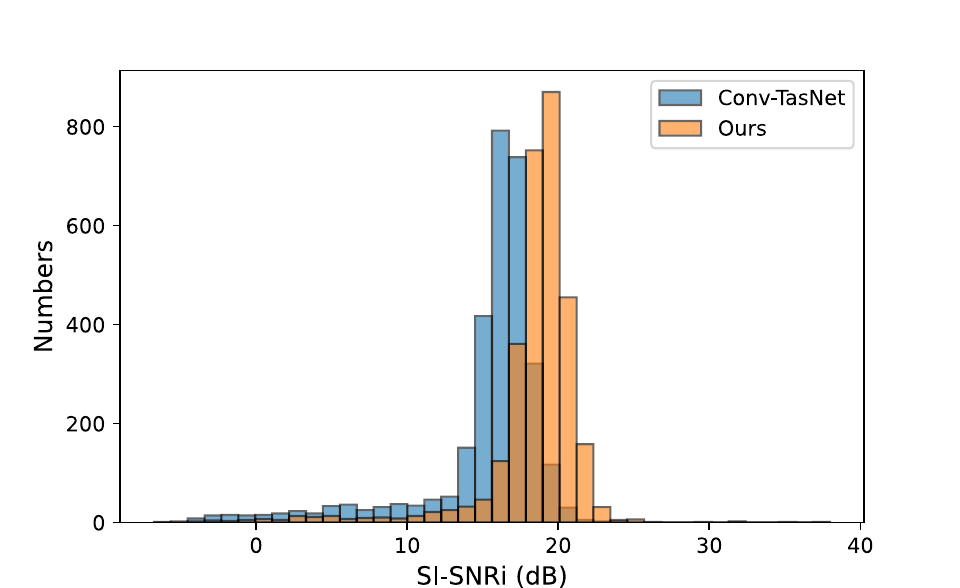}
    \caption{The performance histograms of Conv-TasNet and our eConv-TasNet.}
    \label{fig:per_hist}
\end{figure}
\begin{figure}[t]
    \centering
    \includegraphics[clip, trim= 0.1cm 0cm 0.2cm 0cm, width=\linewidth]{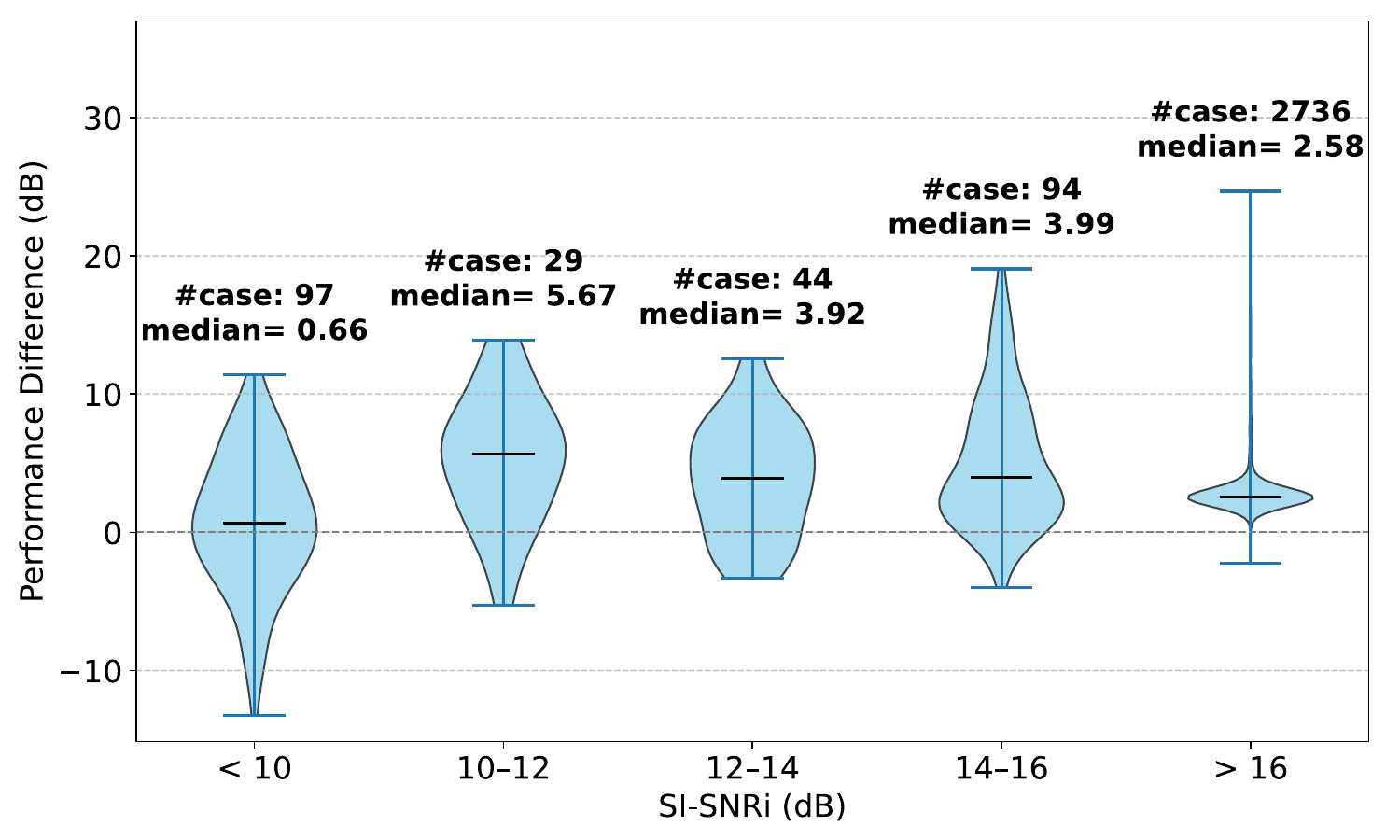}
    \caption{The performance gap between Conv-TasNet and our eConv-TasNet in different performance levels.}
    \label{fig:per_gap}
\end{figure}


\subsection{Performance gain analysis}
To understand the effectiveness of the proposed GES and MGFA modules, we analyze the performance distribution on the WSJ0-2mix test set.
Fig.~\ref{fig:per_hist} presents the SI-SNRi histograms of Conv-TasNet and the proposed eConv-TasNet.
Compared with the baseline, the distribution of eConv-TasNet is shifted toward higher SI-SNRi values, indicating consistent performance improvements across the test set.
In particular, the peak of the distribution moves from approximately 17--18 dB to 19--21 dB, suggesting that a large number of utterances benefit from the proposed architecture.

To further investigate how the improvements vary across different difficulty levels, we group test samples according to the SI-SNRi achieved by Conv-TasNet and reports the performance differences between eConv-TasNet and Conv-TasNet.
As shown in Fig.~\ref{fig:per_gap}, the proposed method consistently outperforms the baseline across all groups.
The largest median improvement of 5.67 dB is observed for instances with SI-SNRi between 10 and 12 dB, followed by gains of 3.99 dB and 3.92 dB for the 14--16 dB and 12--14 dB groups, respectively.
These results indicate that the proposed method is particularly effective for moderately challenging mixtures, where additional speaker-discriminative information can be better exploited.
For extremely difficult cases (SI-SNRi $<$ 10 dB), the median improvement is only 0.66 dB.
This suggests that some mixtures remain difficult to separate even with enhanced speaker representations, reflecting the inherent limitations of TCN-based architectures under highly adverse conditions.
Nevertheless, positive gains are still observed for most samples in this category.

In addition, even for already well-separated mixtures (SI-SNRi $>$ 16 dB), eConv-TasNet achieves a median improvement of 2.58 dB over 2,736 test instances, demonstrating that the proposed modules not only improve challenging cases but also provide consistent refinements for high-quality separations.
Overall, eConv-TasNet outperforms Conv-TasNet on 97\% of the test utterances, demonstrating that the observed gains are broadly distributed across the dataset rather than being driven by a small number of favorable examples.
The results confirm that the proposed architecture provides consistent and robust improvements over Conv-TasNet across a wide range of separation difficulties.

\begin{table}
    \centering
    \caption{Comparison of the distinct configurations in eConv-TasNet on the WSJ0-2mix dataset. The same hyperparameters as the original Conv-TasNet ($L/B/Sc/P$) are skipped in this table.}
    \begin{tabular}{ccccccc}
         \hline
         $N$ & $H$ & $X$ & $R$ & \#Params (M) & SI-SNRi & RTF \\\hline
         256 & 256 & 6 & 3 & 1.6 & 16.1 & 0.06\\ 
         256 & 256 & 6 & 4 & 2.1 & 17.1 & 0.08\\
         256 & 256 & 4 & 6 & 2.1 & 16.3 & 0.09\\
         512 & 512 & 6 & 3 & 3.0 & 17.4 & 0.11\\
         512 & 512 & 8 & 3 & 3.8 & 18.4 & 0.13\\
         512 & 512 & 8 & 4 & 4.9 & 18.8 & 0.17\\
         \hline
    \end{tabular}
    \label{tab:hyperparam}
\end{table}

\begin{table*}[h]
    \centering
    \caption{Comparison with other methods on the WSJ0-2mix, WHAM!, and Libri2Mix datasets.}
    \resizebox{\linewidth}{!}{
    \begin{tabular}{llccccccccc}
        \hline
        \multirow{2}{*}{Methods} & \multirow{2}{*}{Sequence model} & \multirow{2}{*}{\#Params (M)} & \multirow{2}{*}{MACs (G)} & \multicolumn{2}{c}{WSJ0-2mix} & \multicolumn{2}{c}{WHAM!} & \multicolumn{2}{c}{Libri2Mix} & \multirow{2}{*}{RTF} \\\cline{5-10}
        &&&&SI-SNRi & SDRi & SI-SNRi & SDRi& SI-SNRi & SDRi&\\ \hline
        Conv-TasNet~\cite{luo2019conv} & CNN & 4.9 & 10.09 & 15.3 & 15.6 & 12.9 & 13.2 & 12.1 & 12.5 & 0.16\\
        DPRNN~\cite{luo2020dual} & Dual-path + BLSTM & 2.7& 86.31 & 18.8 & 19.0 & 13.7 & 14.1 & 16.1 & 16.6 & 1.18\\
        SuDoRm-Rf~\cite{tzinis2020sudo} & Multi-scale & 2.7 & 4.02 & 17.0 & 17.3 & 12.9 & 13.3 & 13.2 & 13.6 & 0.15\\
        A-FRCNN-4~\cite{hu2021speech} & Multi-scale + Unfolding & 6.1 & 21.19 & 15.6 & 15.8 & 12.0 & 12.3 & - & - & 0.23\\
        A-FRCNN-16~\cite{hu2021speech} & Multi-scale + Unfolding & 6.1 & 80.72 & 18.3  & 18.6 & 14.5 & 14.8 & 16.2 & 16.7 & 0.89\\
        SepFormer~\cite{subakan2021attention} & Dual-path + Transformer & 25.7 & 74.94 & 20.4 & 20.5 & 14.4 & 15.0 & 16.5 & 16.8 & 12.40\\
        TDANet~\cite{li2023efficient} & Multi-scale + Transformer & 2.3 & 9.46 & 18.5 & 18.7 & 15.2 & 15.4 & 17.4 & 17.9 & 0.28\\
        SepReformer-T~\cite{shin2024separate} & Multi-scale + Transformer & 3.9 & 12.56 & 22.4 & 22.6 & 17.2 & 17.5 & 19.7 & 20.2 & 0.68\\
        SepReformer-L~\cite{shin2024separate} & Multi-scale + Transformer& 56.8 & 170.31 & 25.1 & 25.2 & 18.4 & 18.7 & - & - & 3.88\\
        Tiger~\cite{xu2025tiger} & Multi-scale + Transformer & 0.8 & 7.75 & - & - & - & - & 16.7 & 17.1 & 0.85\\ 
        \hline
        Ours (eConv-TasNet)& CNN & 3.8 & 8.86 & 18.4 & 18.6 & 14.7 & 15.0 & 15.6 & 16.0 & 0.13\\
        \hline
    \end{tabular}
    }
    \label{tab:sota}
\end{table*}

\subsection{Performance-efficiency trade-off analysis}
We further investigate the trade-off between separation performance and computational cost using different eConv-TasNet configurations on the WSJ0-2mix dataset.
As shown in Table~\ref{tab:hyperparam}, increasing the network capacity generally improves separation performance at the expense of larger model size and higher inference cost.
Specifically, enlarging the feature channel and hidden layer dimensions ($N$ and $H$) and increasing the number of TCN blocks ($X$ and $R$) consistently improves SI-SNRi, demonstrating the effectiveness of scaling the proposed architecture.
The best model achieves an SI-SNRi of 18.8 dB, representing a 3.5 dB improvement over the baseline Conv-TasNet in Table~\ref{tab:ablation}, while maintaining a comparable model size and inference cost.
Notably, even lightweight configurations with only 1.6M parameters can already outperform the baseline, highlighting the parameter efficiency of the proposed design.
Considering the trade-off between separation quality and computational efficiency, we adopt the configuration with $N=512$, $H=512$, $X=8$, and $R=3$ in the remaining experiments, as it provides a favorable balance between SI-SNRi, model size, and inference cost.




\subsection{Comparison with state-of-the-art methods}
CNN-based architectures, such as Conv-TasNet, are inherently suitable for real-time speech separation due to their fully convolutional design and low-latency inference.
As shown in Table~\ref{tab:sota}, Conv-TasNet achieves significant runtime efficiency; however, it still exhibits suboptimal performance-efficiency trade-offs in terms of model size, MACs, and separation performance compared with more recent architectures incorporating specialized mechanisms.

Recent efficient models, including SuDoRM-RF~\cite{tzinis2020sudo} and A-FRCNN variants~\cite{hu2021speech}, further explore architectural modifications such as multi-scale processing and unfolding mechanisms (i.e., variant U-net and recurrent networks) to improve effectiveness and efficiency.
For more details, SuDoRM-RF achieves relatively low MACs but provides limited performance gains, and the A-FRCNN series demonstrates that improving performance often requires increased model complexity, resulting in less consistent efficiency scaling.
This suggests that efficiency-oriented design alone does not guarantee a balanced improvement in both accuracy and computational cost.
In contrast, state-of-the-art approaches based on dual-path modeling and Transformers, such as DPRNN and SepFormer, achieve strong separation performance, demonstrating the effectiveness of long-range dependency modeling.
However, these improvements come with significantly increased computational overhead and inference latency.
Notably, even compact variants such as TDANet and SepReformer-T, despite relatively lower MACs, still exhibit non-negligible runtime latency (RTF), indicating that MACs alone are not sufficient to characterize real-time feasibility, where inference latency becomes the dominant constraint for edge deployment.

Overall, the results highlight that existing methods typically optimize either efficiency or effectiveness, but rarely achieve both simultaneously under real-time constraints, especially for edge devices without adequate computational resources.
Our eConv-TasNet achieves a more favorable efficiency–effectiveness trade-off by improving separation performance over Conv-TasNet while maintaining low computational cost and competitive real-time inference speed, making it potential for edge and latency-sensitive applications.

\section{Conclusion}\label{sec:conclusion}
In this study, we introduce eConv-TasNet, an efficient speech separation framework that integrates GES and MGFA modules into the original Conv-TasNet architecture to jointly improve both effectiveness and efficiency.
The proposed GES module is designed to compute discriminative speaker representations at intermediate stages of the separator, rather than deferring representation learning to the final stage. 
This design enables earlier and more efficient utilization of speaker information while reducing redundant computations in later layers.
Furthermore, the MGFA module progressively aggregates and refines these speaker-dependent representations, enabling more robust mask estimation through a stepwise fusion mechanism that enhances feature consistency across stages.
Experimental results demonstrate that eConv-TasNet consistently outperforms the original Conv-TasNet across multiple benchmarks, achieving improved separation performance while simultaneously reducing model size and inference latency.
In addition, compared with state-of-the-art methods, eConv-TasNet achieves a more favorable balance between performance and computational cost. 
While many recent approaches improve either separation quality or efficiency, our method provides a potential solution that maintains competitive performance with real-time friendly inference, making it particularly suitable for deployment on resource-constrained edge devices.


\vfill
\pagebreak
\bibliographystyle{IEEEtran}
\bibliography{mybib}

@article{wang2018supervised,
  title={Supervised speech separation based on deep learning: An overview},
  author={Wang, DeLiang and Chen, Jitong},
  journal={IEEE/ACM transactions on audio, speech, and language processing},
  volume={26},
  number={10},
  pages={1702--1726},
  year={2018},
  publisher={IEEE}
}

@article{wang2018end,
  title={End-to-end speech separation with unfolded iterative phase reconstruction},
  author={Wang, Zhong-Qiu and Roux, Jonathan Le and Wang, DeLiang and Hershey, John R},
  journal={arXiv preprint arXiv:1804.10204},
  year={2018}
}

@inproceedings{hershey2016deep,
  title={Deep clustering: Discriminative embeddings for segmentation and separation},
  author={Hershey, John R and Chen, Zhuo and Le Roux, Jonathan and Watanabe, Shinji},
  booktitle={IEEE International Conference on Acoustics, Speech and Signal Processing (ICASSP)},
  pages={31--35},
  year={2016}
}

@inproceedings{yang2023monaural,
  title={Monaural speech separation method based on recurrent attention with parallel branches},
  author={Yang, Xue and Bao, Changchun and Zhang, Xu and Chen, Xianhong},
  booktitle={Interspeech},
  pages={3794--3798},
  year={2023}
}

@inproceedings{li2018cbldnn,
  title={CBLDNN-based speaker-independent speech separation via generative adversarial training},
  author={Li, Chenxing and Zhu, Lei and Xu, Shuang and Gao, Peng and Xu, Bo},
  booktitle={IEEE International Conference on Acoustics, Speech and Signal Processing (ICASSP)},
  pages={711--715},
  year={2018},
}

@inproceedings{luo2018tasnet,
  title={Tasnet: time-domain audio separation network for real-time, single-channel speech separation},
  author={Luo, Yi and Mesgarani, Nima},
  booktitle={IEEE International Conference on Acoustics, Speech and Signal Processing (ICASSP)},
  pages={696--700},
  year={2018},
}

@inproceedings{luo2020dual,
  title={Dual-path rnn: efficient long sequence modeling for time-domain single-channel speech separation},
  author={Luo, Yi and Chen, Zhuo and Yoshioka, Takuya},
  booktitle={IEEE International Conference on Acoustics, Speech and Signal Processing (ICASSP)},
  pages={46--50},
  year={2020},
}

@inproceedings{subakan2021attention,
  title={Attention is all you need in speech separation},
  author={Subakan, Cem and Ravanelli, Mirco and Cornell, Samuele and Bronzi, Mirko and Zhong, Jianyuan},
  booktitle={IEEE International Conference on Acoustics, Speech and Signal Processing (ICASSP)},
  pages={21--25},
  year={2021},
}

@inproceedings{li2023efficient,
  title={An efficient encoder-decoder architecture with top-down attention for speech separation},
  author={Li, Kai and Yang, Runxuan and Hu, Xiaolin},
  booktitle={The International Conference on Learning Representations},
  year={2023},
}

@article{hu2021speech,
  title={Speech separation using an asynchronous fully recurrent convolutional neural network},
  author={Hu, Xiaolin and Li, Kai and Zhang, Weiyi and Luo, Yi and Lemercier, Jean-Marie and Gerkmann, Timo},
  journal={Advances in Neural Information Processing Systems},
  volume={34},
  pages={22509--22522},
  year={2021}
}

@inproceedings{mu2023multi,
  title={Speech separation using an asynchronous fully recurrent convolutional neural network},
  author={Mu, Zhaoxi and Yang, Xinyu and Zhu, Wenjing},
  booktitle={IEEE International Conference on Acoustics, Speech and Signal Processing (ICASSP)},
  year={2023},
}

@inproceedings{zhao2023mossformer,
  title={MossFormer: Pushing the Performance Limit of Monaural Speech Separation Using Gated Single-Head Transformer with Convolution-Augmented Joint Self-Attentions},
  author={Zhao, Shengkui and Ma, Bin},
  booktitle={IEEE International Conference on Acoustics, Speech and Signal Processing (ICASSP)},
  year={2023}
}

@inproceedings{lam2021sandglasset,
  author={Lam, Max W. Y. and Wang, Jun and Su, Dan and Yu, Dong},
  booktitle={IEEE International Conference on Acoustics, Speech and Signal Processing (ICASSP)},
  title={Sandglasset: A Light Multi-Granularity Self-Attentive Network for Time-Domain Speech Separation}, 
  year={2021},
}

@inproceedings{tzinis2020,
  author={Tzinis, Efthymios and Wang, Zhepei and Smaragdis, Paris},
  booktitle={International Workshop on Machine Learning for Signal Processing}, 
  title={Sudo RM -RF: Efficient Networks for Universal Audio Source Separation}, 
  year={2020},
  organization={IEEE}
}

@article{shin2024separate,
  title={Separate and reconstruct: Asymmetric encoder-decoder for speech separation},
  author={Shin, Ui-Hyeop and Lee, Sangyoun and Kim, Taehan and Park, Hyung-Min},
  journal={Advances in Neural Information Processing Systems},
  volume={37},
  pages={52215--52240},
  year={2024}
}

@article{luo2019conv,
  title={Conv-TasNet: Surpassing ideal time-frequency magnitude masking for speech separation},
  author={Luo, Yi and Mesgarani, Nima},
  journal={IEEE/ACM transactions on audio, speech, and language processing},
  volume={27},
  number={8},
  pages={1256--1266},
  year={2019},
  publisher={IEEE}
}

@article{wichern2019wham,
  title={Wham!: Extending speech separation to noisy environments},
  author={Wichern, Gordon and Antognini, Joe and Flynn, Michael and Zhu, Licheng Richard and McQuinn, Emmett and Crow, Dwight and Manilow, Ethan and Roux, Jonathan Le},
  journal={arXiv preprint arXiv:1907.01160},
  year={2019}
}

@article{kolbaek2017multitalker,
  title={Multitalker speech separation with utterance-level permutation invariant training of deep recurrent neural networks},
  author={Kolb{\ae}k, Morten and Yu, Dong and Tan, Zheng-Hua and Jensen, Jesper},
  journal={IEEE/ACM Transactions on Audio, Speech, and Language Processing},
  volume={25},
  number={10},
  pages={1901--1913},
  year={2017},
  publisher={IEEE}
}

@article{zeghidour2021wavesplit,
  title={Wavesplit: End-to-end speech separation by speaker clustering},
  author={Zeghidour, Neil and Grangier, David},
  journal={IEEE/ACM Transactions on Audio, Speech, and Language Processing},
  volume={29},
  pages={2840--2849},
  year={2021},
  publisher={IEEE}
}

@inproceedings{le2019sdr,
  title={SDR--half-baked or well done?},
  author={Le Roux, Jonathan and Wisdom, Scott and Erdogan, Hakan and Hershey, John R},
  booktitle={IEEE International Conference on Acoustics, Speech and Signal Processing (ICASSP)},
  pages={626--630},
  year={2019}
}

@article{vincent2006performance,
  title={Performance measurement in blind audio source separation},
  author={Vincent, Emmanuel and Gribonval, R{\'e}mi and F{\'e}votte, C{\'e}dric},
  journal={IEEE transactions on audio, speech, and language processing},
  volume={14},
  number={4},
  pages={1462--1469},
  year={2006},
  publisher={IEEE}
}

@inproceedings{tzinis2020sudo,
  title={Sudo rm-rf: Efficient networks for universal audio source separation},
  author={Tzinis, Efthymios and Wang, Zhepei and Smaragdis, Paris},
  booktitle={International Workshop on Machine Learning for Signal Processing (MLSP)},
  pages={1--6},
  year={2020},
  organization={IEEE}
}

@article{li2025advances,
  title={Advances in speech separation: Techniques, challenges, and future trends},
  author={Li, Kai and Chen, Guo and Sang, Wendi and Luo, Yi and Chen, Zhuo and Wang, Shuai and He, Shulin and Wang, Zhong-Qiu and Li, Andong and Wu, Zhiyong and others},
  journal={arXiv preprint arXiv:2508.10830},
  year={2025}
}

@article{wang2023tf,
  title={TF-GridNet: Integrating full-and sub-band modeling for speech separation},
  author={Wang, Zhong-Qiu and Cornell, Samuele and Choi, Shukjae and Lee, Younglo and Kim, Byeong-Yeol and Watanabe, Shinji},
  journal={IEEE/ACM Transactions on Audio, Speech, and Language Processing},
  volume={31},
  pages={3221--3236},
  year={2023},
  publisher={IEEE}
}

@article{xu2025tiger,
  title={Tiger: Time-frequency interleaved gain extraction and reconstruction for efficient speech separation},
  author={Xu, Mohan and Li, Kai and Chen, Guo and Hu, Xiaolin},
  journal={The Thirteenth International Conference on Learning Representations},
  year={2025}
}

@inproceedings{li2022skim,
  title={Skim: Skipping memory lstm for low-latency real-time continuous speech separation},
  author={Li, Chenda and Yang, Lei and Wang, Weiqin and Qian, Yanmin},
  booktitle={ICASSP 2022-2022 IEEE International Conference on Acoustics, Speech and Signal Processing (ICASSP)},
  pages={681--685},
  year={2022},
  organization={IEEE}
}

@inproceedings{zhao2024mossformer2,
  title={Mossformer2: Combining transformer and rnn-free recurrent network for enhanced time-domain monaural speech separation},
  author={Zhao, Shengkui and Ma, Yukun and Ni, Chongjia and Zhang, Chong and Wang, Hao and Nguyen, Trung Hieu and Zhou, Kun and Yip, Jia Qi and Ng, Dianwen and Ma, Bin},
  booktitle={ICASSP 2024-2024 IEEE International Conference on Acoustics, Speech and Signal Processing (ICASSP)},
  pages={10356--10360},
  year={2024},
  organization={IEEE}
}

@article{cosentino2020librimix,
  title={Librimix: An open-source dataset for generalizable speech separation},
  author={Cosentino, Joris and Pariente, Manuel and Cornell, Samuele and Deleforge, Antoine and Vincent, Emmanuel},
  journal={arXiv preprint arXiv:2005.11262},
  year={2020}
}

@inproceedings{panayotov2015librispeech,
  title={Librispeech: an asr corpus based on public domain audio books},
  author={Panayotov, Vassil and Chen, Guoguo and Povey, Daniel and Khudanpur, Sanjeev},
  booktitle={2015 IEEE international conference on acoustics, speech and signal processing (ICASSP)},
  pages={5206--5210},
  year={2015},
  organization={IEEE}
}

@article{loshchilov2017decoupled,
  title={Decoupled weight decay regularization},
  author={Loshchilov, Ilya and Hutter, Frank},
  journal={arXiv preprint arXiv:1711.05101},
  year={2017}
}

@techreport{Garofolo1993,
  title       = {CSR-I (WSJ0) Complete LDC93S6A},
  author      = {Garofolo, John S. and Pallett, David S. and Fiscus, Jonathan G. and Fisher, William M. and Hindle, Alvin and Le, David S.},
  institution = {Linguistic Data Consortium},
  address     = {Philadelphia, PA},
  year        = {1993},
  url         = {https://catalog.ldc.upenn.edu/LDC93S6A}
}

@article{bs1770algorithms,
  title={Algorithms to measure audio programme loudness and true-peak audio level},
  author={BS, ITUR and others},
  journal={International Telecommunication Union, Tech. Rep},
  volume={4},
  pages={2015},
  year={1770}
}

@article{agrawal2023review,
  title={A review on speech separation in cocktail party environment: challenges and approaches},
  author={Agrawal, Jharna and Gupta, Manish and Garg, Hitendra},
  journal={Multimedia Tools and Applications},
  volume={82},
  number={20},
  pages={31035--31067},
  year={2023},
  publisher={Springer}
}

@inproceedings{lam2021effective,
  title={Effective low-cost time-domain audio separation using globally attentive locally recurrent networks},
  author={Lam, Max WY and Wang, Jun and Su, Dan and Yu, Dong},
  booktitle={2021 IEEE Spoken Language Technology Workshop (SLT)},
  pages={801--808},
  year={2021},
  organization={IEEE}
}

\end{document}